\documentclass[preprint,superscriptaddress,showpacs,preprintnumbers,amsmath,amssymb]{revtex4}
\usepackage{graphicx}
\usepackage{dcolumn}
\usepackage{bm}

\begin{document}


\title{Radiation-induced correlation between molecules nearby metallic antenna array}

\author{Yoshiki Osaka}
 \email{osaka@pe.osakafu-u.ac.jp}
\author{Nobuhiko Yokoshi}
 \email{yokoshi@pe.osakafu-u.ac.jp}
\author{Hajime Ishihara}
\affiliation{Department of Physics and Electronics, Osaka Prefecture University, 1-1 Gakuen-cho, Sakai, Osaka 599-8531, Japan}

\date{\today}

\begin{abstract}
We theoretically investigate optical absorption of molecules embedded nearby metallic antennas by using discrete dipole approximation method. It is found that the spectral peak of the absorption is shifted due to the radiation-induced correlation between the molecules. The most distinguishing feature of our work is to show that the shift is largely enhanced even when the individual molecules couple with localized surface plasmons near the different antennas. Specifically, we first consider the case that two sets of dimeric gold blocks with a spacing of a few nanometer are arranged, and reveal that the intensity and spectral peak of the optical absorption strongly depends on the position of the molecules. In addition, when the dimeric blocks and the molecules are periodically arranged, the peak shift is found to increase up to $\sim 1.2 {\rm meV} (300 {\rm GHz})$.  Because the radiation-induced correlation is essential for collective photon emission, our result implies the possibility of plasmon-assisted superfluorescence in designed antenna-molecule complex systems.
\end{abstract}

\maketitle

\section{Introduction}
Optically-illuminated metallic nanostructure, such as nanoparticles and nanoblocks, generates an extremely localized electric field due to localized surface plasmon resonance (LSPR), by which we can make individual molecules effectively couple with weak light~\cite{NovotnyQuenching}. In addition, the intensity gradient of the localized field becomes steep enough to break long-wavelength approximation, and excite dipole-forbidden transitions even in a single molecule~\cite{forbiddenTHE,forbiddenTHE2,forbiddenEXP}. These features remind us of sophisticated antenna systems for photons, and have attracted increased attention from various scientific fields with a view to efficient reaction fields~\cite{OpticalAntennas,OpticalAntennas2,OpticalAntennas3}. We have theoretically examined optical responses of antenna-molecule coupled systems, and found that, for certain set of parameters, one can efficiently excite only the molecule with the antenna excitation inhibited (energy transparency effect)~\cite{gapET,nakatani}. Actually, similar effect is experimentally verified using gold nanoantenna system~\cite{gapmode}. Such a transparency is explained by considering the quantum interference between the molecular polarization and the plasmon, and it can be applied also to the nonlinear response by a few photons~\cite{osakaPRL}. 

Besides, it is theoretically proposed that the plasmon-assisted enhancement of light-molecule interaction can be applied to cooperative photoemission such as superradiance and superfluorescence~\cite{PlasmonicDicke,PlasmonicDicke2}. According to the Dicke's theory, the ensemble of emitters can radiate coherent light after creating a macroscopic dipole moment through the incident and radiation fields~\cite{Dicke,SF}. Then, in general, cooperative photoemission requires high-density emitters. Actually, in Ref.~\cite{PlasmonicDicke,PlasmonicDicke2}, all the molecules are assumed to lie under the common localized field. However, if the molecules are efficiently excited by using the energy transparency effect, the radiated field from the molecule should be enlarged. Then, such a plasmon-assisted radiations can enhance inter-molecule correlation.

In the present work, we numerically analyze the optical absorption of molecules near a metallic nanostructure by using discrete dipole approximation (DDA) method~\cite{DDA,DDA2,DDA3}, and investigate the correlation between the molecules. Here it is assumed that the metallic structure consists of dimeric gold nanobloks with a spacing of a few nanometer.  In such a system, the intensity of the localized electric field becomes $\sim 10^5$ times larger than that of the incident light~\cite{bowtie}. In addition, the periodic array of the dimeric blocks is also produced experimentally~\cite{plasmonTPA,arrayMNP}. It is found that the spectral peak of the optical absorption becomes largely shifted even when the molecules couple with different plasmons. We also find that radiation-induced correlation is further enhanced when the dimeric blocks and molecules are periodically arranged. The quantitative study of the inter-molecule correlation is essential to explore the cooperative photoemission in the metallic antenna-molecule coupled system, and may contribute to designing future photoemission devices.  

\section{Model and Method}
\begin{figure}[tb]
\includegraphics[width=80mm]{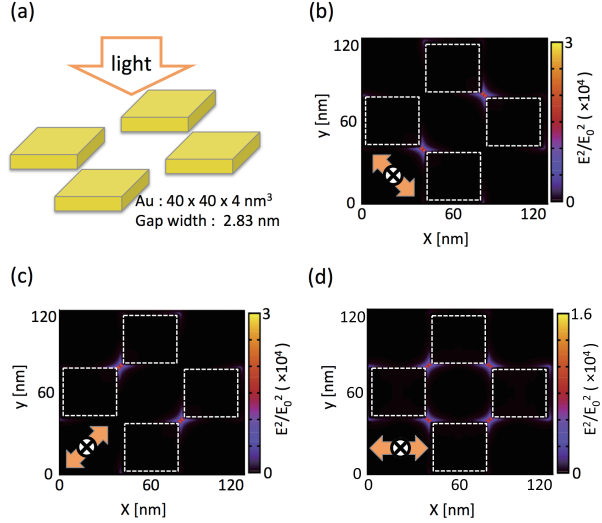}
\caption{(a) Schematic figure of the metallic nanoblocks. We arrange four gold nanoblocks with a small spacing 2.83 nm. The size of each nanoblock is 40 $\times$ 40 $\times$ 4 nm$^3$. (b) The profile of the electric field intensity in the absence of the molecules. The two-way orange arrow indicates the direction of the polarization of the incident field. One can see that the total electric field is strongly enhanced at the spacing, and produces ``hot spots'' for the molecules. (c, d) The profile of the total electric fields is plotted when the direction of the polarization is changed. One can see that the position and the intensity are changed with the direction.}
\label{fig:dda}
\end{figure}
The system under consideration is shown in Fig. \ref{fig:dda}(a). The size of each block is $40 \times 40 \times 4$ nm$^3$, and the spacing between the blocks is 2.83 nm. For a clear demonstration, we have considered the blocks to be sufficiently thin so that the mode volume of the localized light fields becomes comparable to the molecular volume. In order to obtain the optical absorption, we solve the discretized integral form of Maxwell's equation within DDA~\cite{DDA,DDA2,DDA3}. In this calculation, we divide the whole space containing the metal blocks into small cubic cells, where microscopic quantities such as the electric field and polarization in each cell are averaged. The integral equation is given by
\begin{eqnarray}
{\bf E}_i= {\bf E}^0_i+ \sum^n_{j=1} {\bf G}^{f}_{i,j} {\bf P}_j V_j+({\bf M}_i-{\bf L}_i) {\bf P}_i ,
\label{eq.dda}
\end{eqnarray}
where $n$ represents the total number of cells, and $i,j$ are the cell number. The incident field ${\bf E}^0_i$ is linearly polarized along the direction of the orange two-way arrows in Figs. \ref{fig:dda}(b-d), and the polarization in each cell is connected with the total electric field ${\bf E}_i$ as
\begin{eqnarray}
{\bf P}_j =\chi_j {\bf E}_j.
\label{eq.kousei}
\end{eqnarray}
Here, $\chi_j$ and $V_j$ are the optical susceptibility and the volume of the $j$-th cell, respectively. The free space Green's function ${\bf G}^f_{i,j}$ has both transverse and longitudinal electromagnetic components. The term including ${\bf M}_i-{\bf L}_i$ denotes the self-interaction, which is introduced in order to avoid unphysical divergence and can be calculated analytically as~\cite{selfterm1,selfterm2}
\begin{eqnarray}
 {\bf M}_i=\frac{8\pi}{3} {\bf I} \left[ (1-ika) \exp(ika) -1 \right],
\end{eqnarray}
and
\begin{eqnarray}
 {\bf L}_i=\frac{4\pi}{3} {\bf I}.
\end{eqnarray}
Here the parameter $a$ characterize the size of the unit cell, and $k=\omega/c$ is the wave number with $\omega$ being the frequency of the incident field. 

The unit cell in the metal blocks has a Drude-type dielectric function 
\[
\chi_i^{\rm m}= \varepsilon_{\rm bg}^{\rm m}-\varepsilon_{\rm bg}-\frac{(\hbar \omega_p)^2}{[(\hbar\omega)^2+i\hbar\omega(\hbar\gamma_{\rm b}+\hbar\nu_F/D_{\rm eff})]},
\]
where the background dielectric constant of the gold is set to be $\varepsilon_{\rm bg}^{\rm m}=12.0$, and the environmental background dielectric constant to $\varepsilon_{\rm bg}=1.0$, i.e., the dielectric constant of vacuum. As for the other parameters in the gold blocks, we employ the ones in Ref.~\cite{gold1,gold2}: the plasma frequency $\hbar \omega_p=8.958 {\rm eV}$, the nonradiative damping constant $\hbar\gamma_{\rm b}= 72.3 {\rm meV}$, the mean free path of electron $D_{\rm eff}=20 {\rm nm}$, and the Fermi velocity $\nu_F= 1.4\times10^6{\rm m/s}$. As for the susceptibility of the molecule, we assume it to be Lorentzian;
\begin{eqnarray}
\chi_j^m =\frac{(|d|^2/V_j) }{(\hbar \Omega^m-\hbar \omega-i\gamma)},
\end{eqnarray}
where $d=9.7$ Debye is the dipole moment, $\hbar \Omega=1.57$eV is the resonant energy, and $\hbar\gamma=1$meV is the nonradiative damping constant. We have employed these parameters so that the molecular system becomes resonant to LSPR of the gold blocks, which is necessary in order to introduce the energy transparency effect~\cite{gapET,nakatani}.
 
In this work, we mainly calculate the absorption cross section of molecule according to the definition
\begin{eqnarray}
\sigma_{\rm abs}=4\pi \frac{\omega}{c} \int_V d^3{\bf r} \frac{|{\bf E}({\bf r})|^2}{|{\bf E}_0({\bf r})|^2} {\rm Im}\{ \chi({\bf r}) \} 
\label{eq.absorption}.
\end{eqnarray}
It should be noted that the total electric field is derived by numerically solving the simultaneous equations in Eqs. (\ref{eq.dda}, \ref{eq.kousei}), i.e., the electric fields in the metallic antenna and the molecule are self-consistently calculated. We will estimate the correlation between molecules from the spectra of the absorption cross section.

\section{Results} 
Before calculating the optical absorption of the molecules, we show the profile of the electric field intensity for the case where only the four gold blocks are set (see Figs.\ref{fig:dda}(b-d)). One can see that electric field is enormously enhanced nearby the gap structure of the dimeric antenna, so-called ``hot spot''. Therefore, when the molecules are set at the hot spot, the plasmon fields excite the molecules efficiently. Note that the electric field becomes considerably small far from the hot spots, and then we can excite plasmons with small correlations in the distant spots.

\begin{figure}[t]
\includegraphics[width=70mm]{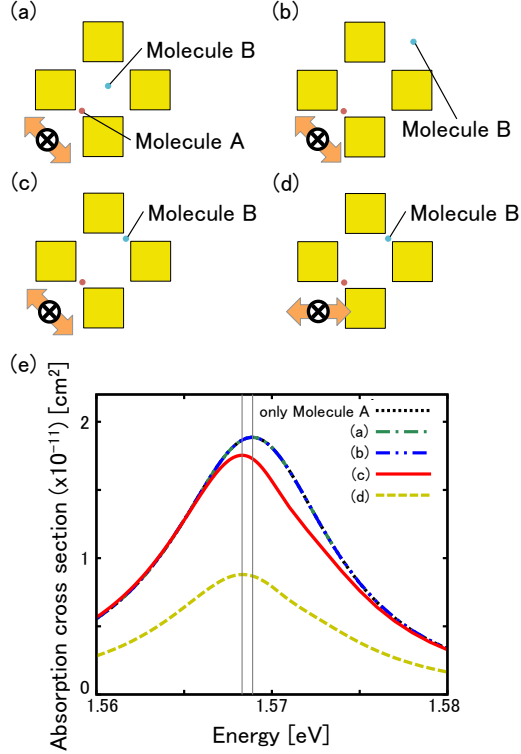}
\caption{(a-d) Top view of the position of two molecules and metal blocks. The polarization direction of the incident field is represented  by two-way orange arrow. (a-c) One of the molecules is fixed (red circle), while the other one (blue circle) is placed in different positions. (d) We alter the polarization direction in the same configuration as (c). (e) Absorption cross-section of the molecule (red circle) is plotted against the incident field energy for different arrangement of molecules. The black dotted line represents the absorption for the case where only one molecule (red circle) exists. One can see the large peak shift ($\sim$0.6meV) due to the inter-molecule interaction, only when both of the molecules are located in the hot spots (Fig. 2 (c,d)).}
\label{fig:nano-gap}
\end{figure}
Subsequently, we investigate the correlation between the two molecules. We arrange the molecules at the position of the red and blue circles in Figs. \ref{fig:nano-gap}(a-d). In Fig. \ref{fig:nano-gap}(e), the calculated absorption cross-sections of the molecule A (red circle) are potted against the energy of the incident light. One can see that, when only one molecule is set in the hot spot, the peak of the absorption is red-shifted by 1.1meV from the resonant energy $\hbar \Omega=1.57$eV (see black dotted line). This shift is caused by the dipole-dipole interaction between the dimeric antenna and the molecule. When another molecule (B) is arranged apart from the hot spot, we cannot see the visible difference in the spectrum (green dashed-dotted line and blue dashed-two dotted line). On the other hand, in case that both of the two molecules are set in different hot spots, the absorption cross section is further red-shifted (red continuous line). We consider that this additional shift originates from the molecule-molecule coupling via the radiation field, because the interference between the plasmons in the distant hot spots has been already included in the case with single molecule. This is the essential difference from the previous work~\cite{PlasmonicDicke,PlasmonicDicke2}. The additional shift reaches 0.6meV (150 GHz) that is comparable to the one by the antenna-molecule coupling. The large inter-molecule correlation appears only when both of the them are excited with high efficiency due to energy transparency effect, and then the plasmon-molecule interference plays the essential role also in the inter-molecule correlation. Besides, when the polarization direction of the incident field is altered, the number of hot spots is also altered. Seeing the yellow dashed line in Fig.~\ref{fig:nano-gap} (e), the  comparable shift is observed even though the absorption cross-section decreases due to the decrease of the effective electric field in one hot spot.

\begin{figure}[t]
\includegraphics[width=80mm]{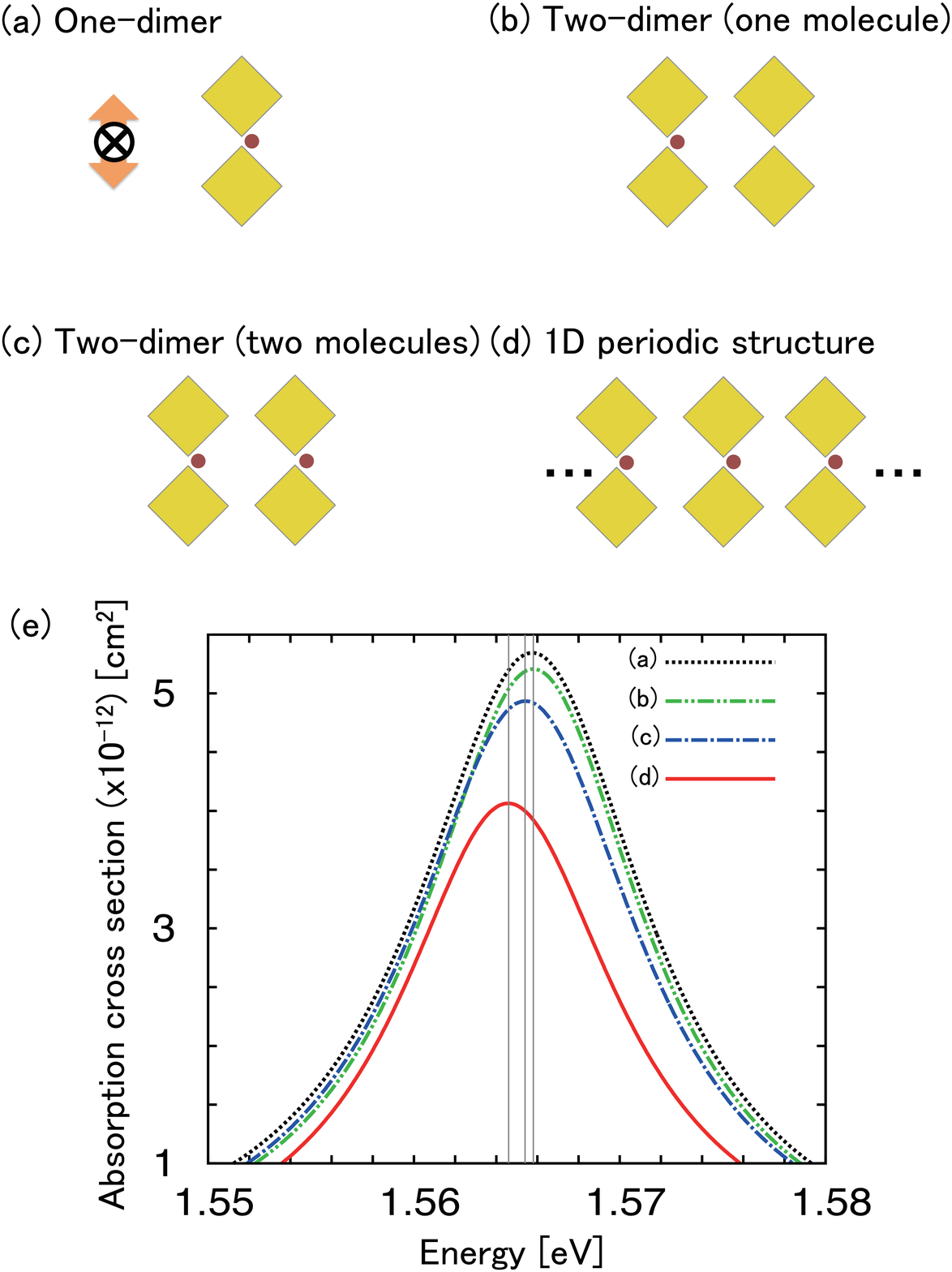}
\caption{(a) Top view of the arrangement of the dimeric antennas and the molecule. Here the spacing between the blocks is changed to be 3nm according to the convenience in the calculation. (b-c) Top view; two sets of the dimeric antennas, and one and two molecules, respectively. (d) Top view of the periodic 1D array of the dimeric antenna-molecule coupled system. (e) The absorption cross-section is plotted for each arrangement pattern. One can see that, in the 1D periodic array, the largest red-shift $\sim $1.2 meV is induced due to the radiation induced inter-molecule correlation, which exceeds the one due to the antenna-molecule coupling.}
\label{fig:array-ab}
\end{figure}
The inter-molecule correlation can be further enhanced by periodically arranging the dimeric antenna-molecule complex. In order to calculate the optical absorption in such a system, we apply the periodic boundary condition to DDA calculation following Ref.~\cite{periodicDDA}. In order to apply the boundary condition, we have changed slightly the spacing between the nanoblocks into 3nm. Figures \ref{fig:array-ab}(a-d) show the arrangement of the metallic antenna and molecules. In Fig. \ref{fig:array-ab}(e), we plot the absorption spectra of one molecule for different arrangements. One can see that the peak shift of the absorption spectrum is no more than 0.4 meV when we add the second set of dimeric antenna and molecule. On the other hand, in case of 1D periodic array structure, the peak shift is increased up to $\sim$1.2 meV (300GHz), which exceeds the one due to dipole-dipole interaction between one dimeric antenna and one molecule.

\section{Conclusion}
With the use of discrete dipole approximation method, we have numerically investigated optical absorption of molecules nearby metallic antennas. We find that large radiation-induced correlation between the molecules can appear even when they couple with different plasmons. The essence of such a large correlation is energy transparency effect due to quantum interference between the plasmon and molecular polarization. Because of the effect, one of the molecules is effectively excited and radiates a photon that couples to the other molecule. We also have evaluated the inter-molecule correlation in periodically arranged antenna-molecule complex, and found that the correlation can be comparable or greater than the single antenna-molecule correlation. 

Although our calculation is based on only linear response theory, the large inter-molecule correlation makes us expect to introduce collective photoemission. Because the metallic antenna can be designed, it may lead to photoemission devices using controlled superfluorescence.  For that aim, we have to calculate optical responses including nonlinear excitations in future work. 

\section{Acknowledgements}
We thank Dr. Y. Mizumoto for helpful discussion on numerical calculation by discrete dipole approximation method. This work was partially supported by a Grant-in-Aid for JSPS Fellows No.13J09308 and  a Grant-in-Aid for Scientific Research (A) No. 24244048 from Japan Society for Promotion of Science (JSPS).

\end{document}